\documentstyle[preprint,tighten,eqsecnum,aps]{revtex}

\begin{document}
\draft
\title{Comment on "Critical currents in ballistic two-dimensional InAs-based superconducting weak
links"}
\author{G.~Bastian}
\address{NTT Basic Research Laboratories, 3-1 Morinosato Wakamiya,
Atsugi, Kanagawa 243-0198, Japan}
\date{\today}
\maketitle

\begin{abstract}

In ballistic Josephson junctions the experimentally observed
iv-characteristics deviate from the theoretically predicted
behaviour. Recently, Heida et al. Phys. Rev. B {\bf 60}, 13135
(1999) discussed this problem and offered an explanation for the
discrepancy. Considering this explanation, several contradictions
to the authors' data as well as to other publications are shown.
\end{abstract}

\narrowtext

In a recent paper, J. P. Heida et al. \cite{Heida99} report on the
iv-characteristics of four ballistic
superconductor-semiconductor-superconductor samples and focus on
possible mechanisms for their observed low characteristic voltage
(the $I_C\times R_N$ product). As many groups struggle with slow
progress to fabricate optimized samples, where this parameter gets
close to the theoretically predicted maximum value, any
explanation why it may not be possible to reach this goal would be
highly welcome. However, the explanation offered rises many
questions, that need to be answered before concluding, that partly
diffusive InAs layers are the key to explain the experimental
findings.

The proper treatment of the InAs layers before the deposition of
the Nb layers has been reported by many groups to be the most
delicate fabrication step in order to achieve low interface
barriers (\cite{Neurohr96} and Refs. 12-20). Using too weak Ar
sputter cleaning, not all of the oxides on the surface are
removed, while too strong cleaning may result in degradation of
the mobility of the sample. Taking a not perfect interface into
account \cite{Comm01}, however, the reduction of the critical
current has been theoretically explained by many authors (e.g.
Refs. 10 and 11).

The first arising question is therefore related to the claim of
perfect interfaces. Evidence for this could be established by
analysis of subgap structures in the iv-characteristics or the
excess current (e.g. using the OBTK-Model as an approximation).
Without such evidence, a reduced critical current in fact is not a
surprise and can be well explained by the established theories. To
show measurements of the characteristics up to voltages of
$2\Delta/e$ might be also helpful in order to extract the real
normal resistance, which for a proper comparison cannot be
approximated by the subgap resistance at low voltages. Especially
the rather low critical temperature of the deposited Nb films for
a thickness of 70~nm seems to indicate the opposite, namely that
some kind of contamination occurred, which most likely makes it
difficult to obtain a high quality interface.

As for the values for the critical current $I_C$ measured for four
different samples, I would appreciate to see this behaviour in a
greater variety of samples before concluding that $I_C$ is
"independent of the junction length". There seems to be too large
parameter spread to support such conclusion, especially when
taking into account more data previously published by the same
authors, that were measured at very similar samples
\cite{Heida98}. According to the authors' explanation, any
parameter spread (in particular true for sample "B" or "4") must
be due to local differences in the properties of the diffusive
InAs layer, as all interfaces are assumed to be ideal. Therefore
the authors should (first) explain this unexpected high parameter
spread. Taking into account the proper effective masses at the
carrier concentration used \cite{Fuchs94}, the coherence length is
found to be about 250~nm and not 500~nm as claimed by the authors.
Therefore none of the samples belongs to the short junction regime
contrary to the authors' claim. This again shows the contradiction
between the claimed result, that all samples are short junctions
and might need to be further discussed.

The most important question, however, concerns the mechanism for
the reduction of the critical current due to scattering events at
the damaged InAs underneath the Nb layers. About 30 modes are
assumed to be formed within the width of 700~nm and scattering
between the modes is claimed to reduce the critical current. The
scattering events however are not necessarily phase destructive,
which is the mechanism to destroy Andreev bound states and to
reduce the critical current. In order to support the authors'
idea, a comparison of samples with different width W and
correspondingly different number of modes would be helpful to
check the proposed $1/N$ behaviour. Samples studied by other
groups (\cite{Bastian99} and references 12-20) with much larger
widths show in contradiction to the authors' explanation higher
characteristic voltages in spite of the much larger number of
modes. In addition, the authors only give an explanation for a
reduced critical current and not for a reduced characteristic
voltage.

Finally, the authors make one claim, which is not in accordance
with recent publications. The authors should state their
arguments: To my knowledge, it has never been shown, that the
pairing interaction inside the normal conducting layer is (always)
absent. As explained in the very first publications about this
kind of Josephson junctions (e.g. Ref. 2), this is only a
simplified assumption, used in most theoretical descriptions
(except e.g. \cite{Datta96,Jacobs99}). On the contrary, many
experimental results indicate the opposite
(\cite{Nitta94,Bastian98} and Ref. 20), as the phonon-mediated
interaction might not end abruptly at any interface.

In conclusion, more data are needed and obvious contradictions
have to be discussed to find a correct description for this kind
of ballistic Josephson junctions.

\end{document}